\documentclass[pra, 12 pt]{revtex4}
\usepackage[english]{babel}
\usepackage{amsmath}
\usepackage{epsfig}

\begin{document}

\title{\bf Theory of degenerate Bose gas without anomalous
averages}
\author{V.B. Bobrov$^{1}$, S.A. Trigger$^{1,\,2}$}
\address{$^1$ Joint\, Institute\, for\, High\, Temperatures, Russian\, Academy\,
of\, Sciences, 13/19, Izhorskaia Str., Moscow\, 125412,
Russia;\\$^2$ Eindhoven  University of Technology, P.O. Box 513,
MB 5600 Eindhoven, The Netherlands\\ email:\,satron@mail.ru}

\begin{abstract}
Theory of a weakly non-ideal Bose gas in the canonical ensemble is
developed without assumption of the C-number representation of the
creation and annihilation operators with zero momentum. It is
shown that the pole of the "density-density"\, Green's function
exactly coincides with the Bogolybov's phonon-roton spectrum of
excitations. At the same time, a gap exists in the one-particle excitation
spectrum . This gap is related to the
density of particles in the "condensate".\\

PACS number(s): 05.30.Jp, 03.75.Kk, 03.75.Nt, 05.70.Fh\\

\end{abstract}

\maketitle

\section{Introduction}

Starting with the Bogolyubov's papers [1,2], the microscopic
theory of the degenerate Bose gas has been based on the special
assumption that the creation $a_0^{+}$ and annihilation $a_0$
operators of particles with zero momentum can be replaced by a
C-number
\begin{eqnarray}
a_0^{+}=a_0= (<N_0>)^{1/2}, \label{C1}
\end{eqnarray}
where $<N_0>$ is the average quantity of particles in the state
with ${\bf p}=0$ ("condensate").

This assumption leads to the necessity of introducing in the theory
the \emph{anomalous averages} ("quasi-averages"), which is
unnatural for the homogeneous and isotropic system under
consideration. Let us pay attention that statement (1) in Bogolyubov's papers is directly related to the requirement of the weak interparticle interaction in the degenerated Bose gas. Though the Bogolyubov theory (see, e.g., [3,4])
gives rise to many important and widely recognized results, such
as the expression for the spectrum of excitations, explanation of
the experimental data and agreement with the Landau superfluidity
condition, there are serious doubts on the validity of the
relation as well as on the agreement of the Hamiltonian
corresponding to the assumption (\ref{C1}) with the original
Hamiltonian of the system under consideration [5,6].

In this work, we propose a selfconsistent\, noncontradictory
description of the degenerate Bose gas in which we do not use the
assumption (\ref{C1}). In this way we show, in particular, that there exists
the gap in the spectrum of one-particle excitations.
The statement that the spectrum of one-particle excitations has a
gap in parallel with the usual phonon-roton branch of the
collective excitations in the degenerate Bose gas has been formulated
in [7].  Later, the existence of the gap has been suggested in [8].
Spectra, thermodynamics, and dynamical structure factor for weakly
non-ideal Bose gas below the condensation temperature $T<T_0$ have
been considered in [9-11] in terms of the dielectric formalism,
similar to that in plasma-like systems. The dielectric formalism based on
the Bogolyubov assumption (\ref{C1}) for the
operators $a_0^{+}$ and $a_0$ was developed in [12]. Recently, the
gap-existence problem for the weakly non-ideal Bose gas has been
considered by different methods in papers [13-16]. In fact, the
paper [14] confirmed (independently) and developed not only the
statement of Ref. [7] about the existence of a one-particle gap in the
weakly non-ideal Bose gas below the condensation temperature, but
also the results of the papers [9-11] on the possibility of
considering the superfluid system without symmetry breaking. Surely, the existence of the gap disagrees with the
conventional opinion
that the gap is missing [3]. But we do believe that the conclusion
about a missing gap is related to the Bogolubov simplifying
assumption (\ref{C1}), and is not obliged to persist in a more
rigorous theory.

It is necessary to stress that in the previous papers both Landau and
Bogolyubov admitted that the one-particle excitation spectrum can
have a gap, but later they dropped this idea, because of
contradiction with observations of the phonon-roton branch of
spectrum at small $q$ in the neutron scattering experiments. The
approach which includes the both spectra, with and without a gap,
has not been found and coexistence of these two branches of
excitations was not suggested.

In the present work, we suggest that in the limit of strong degeneracy $T\rightarrow
0$, where $T$ is the system temperature, all the particles
tend to occupy the zero-momentum state, which means that
\begin{eqnarray}
\lim_{T\rightarrow 0}<N_0>=<N>, \label{C2}
\end{eqnarray}
where $<N>$ is the average number of particles in the system. This
statement is confirmed in the paper by the self-consistent
consideration. Below we use the canonical ensemble where $N$ is a
given C-number.

\section{Statistical sum and averages in canonical ensemble}

Let us consider the Hamiltonian of the non-ideal Bose gas of
particles with zero spin and a mass $m$ in a volume $V$
\begin{eqnarray}
H=\sum_p \varepsilon_{\bf p}\, a^+_{{\bf p}} a_{{\bf
p}}+\frac{1}{2V}\sum_{{\bf q, p_1,p_2}} u(q)a^+_{{\bf
p_1-q/2}}a^+_{{\bf p_2+q/2}}a_{{\bf p_2-q/2}}a_{{\bf p_1+q/2}},
\label{C3}
\end{eqnarray}
where $a^+_{p}$ and $a_{p}$ are the creation and annihilation
operators of particles with the momentum $\hbar{\bf p}$,
\begin{eqnarray}
[a_{{\bf p_2}}, a^+_{{\bf p_1}}]=a_{{\bf p_2}} a^+_{{\bf
p_1}}-a^+_{{\bf p_1}} a_{{\bf p_2}}=\delta_{{\bf p_1},{\bf p_2}},
\label{C4}
\end{eqnarray}
$\varepsilon_{\bf p}=\hbar^2{\bf p}^2/2m$ is the energy
spectrum of a free particle and $u(q)$ is the Fourier-component of
the inter-particle interaction potential.

It is convenient to extract the particular term $U_0$ with ${\bf
q}=0$ from the sum over ${\bf q}$ in the Hamiltonian (\ref{C3}).
This term can be written as
\begin{eqnarray}
U_0=u(0)\frac{N(N-1)}{2V}, \,\; N=\sum_p a^+_{{\bf p}} a_{{\bf
p}}, \;\,\;u(0)=u(q=0)=u(q\rightarrow 0),\label{C5}
\end{eqnarray}
where $N$  is the operator of the total number of particles, and $N$
is the C-number $N=<N>$. Hereafter, the brackets $<...>$ mean
the canonical-ensemble averaging. Since $U_0$ is also the C-number,
one can write the statistical sum of the system under
consideration in the form
\begin{eqnarray}
Z=Sp\, \exp(-H/T)=\exp\left(-u(0)\frac{N(N-1)}{2VT}\right)Sp\,
\exp(-H_0/T),\label{C6}
\end{eqnarray}
where
\begin{eqnarray}
H_0=H-U_0=\sum_p \varepsilon_p\, a^+_{{\bf p}} a_{{\bf
p}}+\frac{1}{2V}\sum_{{\bf q\neq 0, p_1,p_2}} u(q)a^+_{{\bf
p_1-q/2}}a^+_{{\bf p_2+q/2}}a_{{\bf p_2-q/2}}a_{{\bf p_1+q/2}},
\label{C7}
\end{eqnarray}
Therefore, to provide convergence of the statistical sum
(\ref{C7}) in the thermodynamic limit $V\rightarrow \infty$ $N
\rightarrow \infty$, $n=N/V=const.$, the known condition $u(0)> 0$
has to be fulfilled. As $U_0$ is a C-number, it does not affect
any averaging at all, and for calculating of average values the
Hamiltonian $H$ is equivalent to $H_0$. For an arbitrary operator
$<A>$, we get
\begin{eqnarray}
<A>=Z^{-1} Sp\, \{\exp(-H/T)A \}=Z_0^{-1}Sp \,\{\exp(-H_0/T)A
\}\equiv <A>_0; \;\,\;  Z_0=Sp\, \exp(-H_0/T)\label{C8}
\end{eqnarray}

Let us now consider a more complicated situation associated with
calculation of a time dependent correlation functions $f(t)$ such as
\begin{eqnarray}
f(t)=<[A(t), B(0)]>, \;\;\; A(t)=\exp(i H t/\hbar)A\exp(-i H
t/\hbar). \label{C9}
\end{eqnarray}
The time dependence of the operators $a^+_p(t)$ and $a_p(t)$ can
be represented in the form
\begin{eqnarray}
a^+_p(t)=\exp(iH_0 t/\hbar) a^+_p \exp(-iH_0t/\hbar)\exp(i N u (0)
t/V \hbar). \label{C10}
\end{eqnarray}
\begin{eqnarray}
a_p(t)=\exp(-i N u(0)t/V \hbar) \exp(iH_0 t/\hbar) a_p
\exp(-iH_0t/\hbar). \label{C11}
\end{eqnarray}
Therefore, if in each of the operators $A$ and $B$ the numbers of
creation and annihilation operators coincide (which is typical of
the operators of the physical variables), the time-dependent
correlation function can be written as
\begin{eqnarray}
f(t)=<[A(t), B(0)]>_0, \;\;\; A(t)=\exp(i H_0 t/\hbar)A\exp(-i H_0
t/\hbar). \label{C12}
\end{eqnarray}
On this basis, in what follows,
we consider the average values with the Hamiltonian $H_0$
(\ref{C7}) within the canonical ensemble. The free energy of the initial system with the
Hamiltonian $H$ (\ref{C3}), according to (\ref{C5})-(\ref{C7}),
reads
\begin{eqnarray}
F=-T \ln Z=U_0+F_0, \;\;\; F_0=-T \ln Z_0. \label{C13}
\end{eqnarray}

\section{Equations for "density-density"\, Green function}

Experimentally, the spectrum of collective excitations is found
usually from data on the well observable maxima in the dynamical
structure factor $S({\bf q},\omega)$ for ${\bf q} \neq 0$,
\begin{eqnarray}
S({\bf q},\omega)=\frac{1}{V}\int^\infty_{-\infty} \,exp(i\omega
t)<\rho_{{\bf q}} (t)\rho_{-{\bf q}} (0)>_0 dt, \label{C14}
\end{eqnarray}
\begin{eqnarray}
\rho_{{\bf q}} (t)= \sum_p \, a^{+}_{{\bf p-q}/2}(t)a_{{\bf
p+q}/2}(t), \label{C15}
\end{eqnarray}
where $\rho_{{\bf q}} (t)$ is the Fourier-component of the density
operator in the Heisenberg representation. The
dynamical structure factor $S({{\bf q}},\omega)$ (\ref{C14}) is
directly related to [17] with the retarded density-density Green's
function $\chi^R({{\bf q}},z)$ which is analytical in the upper
semi-plane of the complex variable $z$ ($Im z>0$),
\begin{eqnarray}
S({{\bf q}},\omega)=-\frac{2\hbar}{1- \exp(-\hbar \omega/T)}\,Im
\chi^R({{\bf q}},\omega+i0), \label{C16}
\end{eqnarray}
\begin{eqnarray}
\chi^R({{\bf q}},z)=-\frac{i}{\hbar V}\int^\infty_0 dt \,\exp(i z
t)<[ \rho_{{\bf q}} (t)\rho_{-{\bf q}}
(0)]>_0=\frac{1}{V}<<\rho_{{\bf q}} \mid \rho_{-{\bf q}}>>_{z},
\label{C17}
\end{eqnarray}
The definitions (\ref{C16}),(\ref{C17}) have to be taken in the
thermodynamic limit, where
\begin{eqnarray}
\lim_{T\rightarrow 0} <N_0>_0=<N>_0=N. \label{C18}
\end{eqnarray}

According to Eq.~(\ref{C14}), the retarded function $\chi^R(q,z)$ can be
represented in the form
\begin{eqnarray}
\chi^R ({{\bf q}},z) = \frac{1}{V} \,\sum_p \,F({\bf p,q},z),\,\;
F({\bf p,q},z)= \langle\langle \, a^{+}_{{\bf p-q}/2}a_{{\bf
p+q}/2} \mid \rho_{-q} \rangle\rangle_{z} \label{C19}
\end{eqnarray}
The equation of motion for the function $F({\bf p,q},z)$ with the
Hamiltonian $H_0$, determined by (Eq.~(\ref{C7})), can be written
in the form
\begin{eqnarray}
\left(\hbar z +\varepsilon_{\bf{p-q}/2}-\varepsilon_{\bf{p+q}/2}\right)\,F({\bf p,q},z)= f_{{\bf p-q}/2}-f_{{\bf p+q}/2}-\nonumber\\
\frac{1}{V}\sum_k u(k)\sum_{p_1}\langle\langle \, (a^{+}_{{\bf
p+k-q}/2}a^{+}_{{\bf p_1-k}/2}a_{{\bf p_1+k/2}}a_{{\bf p+q}/2}-
a^{+}_{{\bf p-q}/2}a^{+}_{{\bf p_1-k}/2}a_{{\bf p_1+k}/2}a_{{\bf
p-k+q}/2} )\mid \rho_{-q} \rangle\rangle_{z} \label{C20}
\end{eqnarray}
Here $f_{\bf p}$ is the one-particle distribution function over the momenta
$\hbar {\bf p}$
\begin{eqnarray}
f_{\bf p}=<a^{+}_{{\bf p}}a_{{\bf p}}>_0 \label{C21}
\end{eqnarray}
For a temperatures $T<T_0$, where $T_0$ is the condensation temperature,
the one-particle distribution function $f_{\bf p}$
can be represented as [18]
\begin{eqnarray}
f_{{\bf p}}=<N_0>\delta_{{\bf p},0}+f^T_{{\bf p}}(1-\delta_{{\bf
p},0}), \label{C22}
\end{eqnarray}
where \;$N_0=a_0^{+}a_0$ is the operator of the number of
particles with zero momentum ("condensate")\;,
\;\;\;\;$f^T_{{\bf p}}=<a^{+}_{{\bf p}}a_{{\bf p}}>_0$ is the
one-particle distribution function with non-zero momenta (the
"overcondensate"\, states). Therefore,
\begin{eqnarray}
n=n_0 +\frac {1}{V} \sum_{p \neq 0} f^T_{{\bf p}}=n_0 + \int
\frac{d^3 p}{(2\pi)^3} f^T_{{\bf p}}, \label{C23}
\end{eqnarray}
where $n_0=<N_0>/V$ is the average density of particles in the
condensate. From (\ref{C20}) and (\ref{C22}), we find that the
function $F({\bf p,q},z)$ has singularities at ${\bf p}=\pm {\bf
q}/2$. Therefore, the density-density function \, $\chi^R(q,z)$
(\ref{C19}) can be written as
\begin{eqnarray}
\chi^R (q,z) = \frac{1}{V}\,F({\bf q}/2,{\bf
q},z)\,+\frac{1}{V}\,F(-{\bf q}/2,{\bf q},z)\,+
\frac{1}{V}\sum_{p\neq\pm q/2}\,F^T({\bf p,q},z)\label{C24}
\end{eqnarray}
The index $T$ means, that the respective function describes the
"overcondensate"\, particles. The singularities (conditioned by
the condensate) in this function are absent. Then in the last term
in Eq.~(\ref{C24}), we can change summation by integration over
momenta. The functions $F(\pm {\bf q/2,q},z)$, extracted above,
satisfy, according to Eqs.~(\ref{C20}),(\ref{C22}), to the exact
equations of motion
\begin{eqnarray}
\left(\hbar z -\varepsilon_q\right)\,F({\bf q}/2,{\bf
q},z)=[\langle N_0 \rangle-f^T_{{\bf q}}]-\frac{1}{V}\sum_{k\neq
0} u(k)\sum_{p_1}\langle\langle \, (a^{+}_{\bf k} a^{+}_{{\bf p_1}-{\bf k}/2}a_{{\bf p_1}+{\bf k}/2}a_{\bf q} -\nonumber\\
a^{+}_0 a^{+}_{{\bf p_1}-{\bf k}/2}a_{{\bf p_1}+{\bf k}/2}a_{{\bf
q-k}})\mid \rho_{-q} \rangle\rangle_{z} \label{C25}
\end{eqnarray}
\begin{eqnarray}
\left(\hbar z+\varepsilon_q\right)\,F(-{\bf q}/2,{\bf
q},z)=-[\langle N_0 \rangle-f^T_{{\bf q}}]-\frac{1}{V}\sum_{k\neq
0}u(k)\sum_{p_1}\langle\langle \, (a^{+}_{{\bf k-q}} a^{+}_{{\bf p_1}-{\bf k}/2}a_{{\bf p_1}+{\bf k}/2}a_0 -\nonumber\\
a^{+}_{-{\bf q}} a^{+}_{{\bf p_1}-{\bf k}/2}a_{{\bf p_1}+{\bf
k}/2}a_{-{\bf k}})\mid \rho_{-q} \rangle\rangle_{z}. \label{C26}
\end{eqnarray}
Let us further consider the case of strongly degenerate gas, where
$T\rightarrow 0$. To find for this case the main terms, following
to the Bogolyubov's procedure, let us extract the terms on the right-hand sides
of Eqs.~(\ref{C25}),(\ref{C26}), which are determined by
the maximum number of the operators $a^{+}_0$ and $a_0$. In the
limit of strong degeneration, taking into account(\ref{C18}), we
can omit the other terms in the course of calculation of $F$.
Then, since $f^T_{{\bf q}}=f^T_{-{\bf q}}$ (${\bf q}\neq 0$),
Eqs.~(\ref{C25}),(\ref{C26}) take the form
\begin{eqnarray}
\left(\hbar z \mp \varepsilon_q\right)\,F^{(0)}(\pm {\bf q}/2,{\bf
q},z)=\pm [\langle N_0 \rangle-f^T_{{\bf q}}]\pm \frac{1}{V} u(q)
\langle\langle \, (a^{+}_0 a^{+}_0 a_{\bf q} a_0 + a^{+}_0
a^{+}_{-{\bf q}}a_0 a_0)\mid \rho_{-q} \rangle\rangle_{z}
\label{C27}
\end{eqnarray}
Eqs. (\ref{C27}) are exact at low temperature, since they
define the functions $F^{(0)}(\pm {\bf q}/2,{\bf q},z)$. For
calculation the Green's functions in the right side of Eqs.
(\ref{C27}), it is necessary to make some approximations.

\section{Determination of the collective excitations}

It should be emphasized that the Bogolyubov's approach, in which
the operators $a^{+}_0$ and $a_0$ are considered as C-numbers,
leads to violation of the exact relations (\ref{C25}) and
(\ref{C26}). In this approach (\ref{C1}), the main term $F^{(0)}$
of the function $F$ has the form [19]
\begin{eqnarray}
F^{(0)}(\pm {\bf q}/2,{\bf q},z)= \langle N_0 \rangle\,
\langle\langle \,a_{\pm\bf q}\mid a^{+}_{\pm \bf q}
\rangle\rangle_{z} \label{C28}
\end{eqnarray}
Respectively, instead of the exact equations for two-particle Green's
functions (\ref{C25}) and (\ref{C26}), we obtain the equation of
motion for one-particle Green's functions form (\ref{C28}). As is
shown below, the equations of motion for the two-particle Green's
functions without the approximation on the C-number representation
of the operators $a^{+}_0$ è $a_0$ are essentially different from
the equations for the one-particle distribution functions.
Therefore, for calculation of the Green's functions in the right
part of (\ref{C27}), the assumption (\ref{C1}) cannot be used. At
the same time, the idea of the C-number approximation for some
operators seems, itself, very attractive. Below we apply this idea to
the varsion alternative to the Bogolyubov assumption. According to
(\ref{C18}), it is natural to accept that to calculate the
Green's functions in the limit of strong degeneracy $T \rightarrow
0$, the C-number approximation has to be applied not to the
operators $a^{+}_0$ è $a_0$, but to the physical value - the operator of the number of
particles in the "condensate"\, $N_0$
\begin{eqnarray}
N_0= \langle N_0 \rangle. \label{C29}
\end{eqnarray}
For low temperatures $T$, the operator $N_0$ in the averages (under the angle brackets) can be changed by the total density operator $N$ (equal to $\rho_{q=0}$), which is the C-number in the canonical ensemble. Therefore, in the case under consideration, it can be removed from brackets. Therefore, for low temperatures, the proposed approximation seems asymptotically exact.

In this approximation from (\ref{C27}) directly follows
\begin{eqnarray}
\left(\hbar z - \varepsilon_q\right)\,F^{(0)}( {\bf q}/2,{\bf
q},z)=[\langle N_0 \rangle-f^T_{{\bf q}}]+\frac{<N_0>}{V}
u(q)\{F^{(0)}({\bf q}/2,{\bf q},z)+ F^{(0)}(-{\bf q}/2,{\bf
q},z)\} \label{C30}
\end{eqnarray}

\begin{eqnarray}
\left(\hbar z + \varepsilon_q\right)\,F^{(0)}(- {\bf q}/2,{\bf
q},z)=- [\langle N_0 \rangle-f^T_{{\bf q}}]-\frac{<N_0>}{V}
u(q)\{F^{(0)}({\bf q}/2,{\bf q},z)+F^{(0)}(-{\bf q}/2,{\bf q},z)\}
\label{C31}
\end{eqnarray}

From Eqs.~(\ref{C30}),(\ref{C31}), one can find the solutions for
the functions $F(\pm q/2,q,z)$
\begin{eqnarray} F({\bf q}/2,{\bf
q},\hbar z)=\frac{[<N_0>-f^T_{{\bf q}}](\hbar
z+\varepsilon_q)}{(\hbar z)^2-(\hbar\omega(q))^2};\;\; F(-{\bf
q}/2,{\bf q},z)=-\frac{[<N_0>-f^T_{{\bf q}}](\hbar
z-\varepsilon_q)}{(\hbar z)^2-(\hbar\omega(q))^2}\label{C32}
\end{eqnarray}
\begin{eqnarray}
\hbar\omega(q)\equiv \sqrt {\varepsilon_q^2+2 n_0 u(q)
\varepsilon_q}\label{C33}
\end{eqnarray}
The relation (\ref{C33}) for the spectrum $\hbar\omega(q)$
corresponds exactly to the known Bogolyubov expression [1,2]. By
substituting to (\ref{C19}), and taking into account that for the
case of strong degeneration the contribution of the functions
$F^T({\bf p,q},z)$ is negligible, we obtain the expression for the
main term $\chi^{(0)}(q,z)$ of the "density-density"\, Green
function $\chi(q,z)$
\begin{eqnarray}
\chi^{(0)}(q,z) = \frac{2 n_0 \varepsilon_q}{(\hbar
z)^2-(\hbar\omega(q))^2}\left\{1-\frac{f_q^T}{<N_0>}\right\}\label{C34}
\end{eqnarray}
As it is well known, the singularities of the function
$\chi^R(q,z)$ determine the spectrum of collective excitations in
the system. Therefore, under the assumption about C-number
behavior of the operator $N_0$, we obtain the Bogolyubov's result
for the spectrum of the collective excitations in the degenerated
and weakly interacting Bose gas. However, the question on the braced term
$f_{{\bf q}}^T/<N_0>$ in (\ref{C34}) remains open. The problem is in the behavior of the function $f_{{\bf
q}}^{id}$ for the ideal Bose gas [20]
\begin{eqnarray}
f^{id}_{{\bf q}}=\left\{\exp\left(\frac{\varepsilon(
q)}{T}\right)-1\right\}^{-1} \label{C35}
\end{eqnarray}
In the limit of small wave vectors $q$ the function $f^{id}_{{\bf
q}}$ converges at non-zero temperatures (hereafter,
$1/q^2$-divergence). Moreover,
\begin{eqnarray}
\lim_{T\rightarrow 0}\lim_{q \rightarrow 0}f^{id}_q\neq
\lim_{q\rightarrow 0}\lim_{T \rightarrow 0}f^{id}_q \label{C36}
\end{eqnarray}
The similar problem arises when (\ref{C22}) is used.

\section{One-particle excitations}

To calculate the distribution function $f_q^T$ for the
"overcondensate"\, particles in Bose gas, let us consider the
one-particle Green function $g^R(q,z)$
\begin{eqnarray}
g^{R}(q,z) = \langle\langle \, a_{\bf q} \mid a^{+}_{\bf q}
\rangle\rangle_{z},\;\; {\bf q} \neq 0\label{C37}
\end{eqnarray}
This function is directly connected with the distribution function
$f_q^T$ by the relation [21]
\begin{eqnarray}
f_q^T=\int_{-\infty}^\infty
\frac{d\omega}{2\pi}g^{<}(q,\omega),\;\; g^{<}(q,\omega)= -2\hbar
\left\{\exp\left(\frac{\hbar\omega}{T}\right)-1\right\}^{-1} Im\,
g^{R}(q,\omega+i0)\label{C38}
\end{eqnarray}
The equation of motion for the Green function $g^{R}(q,z)$ for $q
\neq 0$ reads
\begin{eqnarray}
\left(\hbar z - \varepsilon_q\right)\,g^{R}(q,z)=1+
\frac{1}{V}\sum_{k\neq 0} u(k)\sum_{p}\langle\langle \,
a^{+}_{{\bf p+k}} a_{{\bf p}}a_{{\bf q+k}}\mid a^{+}_{\bf q}
\rangle\rangle_{z}. \label{C39}
\end{eqnarray}
As for the "density-density"\, Green function, we consider the
case of a strong degeneration and extract
 the terms with maximum quantity of the
operators $a^{+}_0$ and $a_0$ in the right-hand side of Eq.~(\ref{C39}).  Then from Eq.~(\ref{C39}), we find
\begin{eqnarray}
\left(\hbar z - \varepsilon_q\right)\,g^{R}(q,z)=1+\frac{1}{V}
u(q)\left\{\langle\langle \, a^{+}_0 a_{\bf q} a_0 \mid a^{+}_{\bf q}
\rangle\rangle_{z}+\langle\langle \, a^{+}_{-\bf q} a_0 a_0 \mid a^{+}_{\bf q}
\rangle\rangle_{z}\right\}(1-\delta_{q,\,0}), \label{C40}
\end{eqnarray}
In this case, it is assumed that the separated main term on the right-hand side of (40) corresponds to the case of the weak interaction. In the general case, when the interparticle interaction potential $u(q)$ is not small, in separating the terms with a maximum number of the operators $a^{+}_0$ and $a_0$ on the right-hand side of (40), it is necessary to write an infinite series on the perturbation theory on the interaction $u(q)$. It is clear that a similar situation occurs when using (27) instead of (26). Thereby, strictly speaking, the question whether the terms separated on the right-hand side of (40) are main from the viewpoint of the passage to the limit of the weak interaction remains open.

If we now apply the assumption (1) about the C-number representation of particle creation $a^{+}_0$ and annihilation $a_0$ operators at a zero momentum to the calculation of Green's functions on the right-hand side of (40), we come to the necessity of introducing the so-called "anomalous" Green's functions
\begin{eqnarray}
g_{anom}^{R}(q,z)=
\langle\langle \, a^{+}_{-\bf q}  \mid a^{+}_{\bf q}
\rangle\rangle_{z}, \qquad {\bf q}\neq 0, \label{C41}
\end{eqnarray}
Let us pay attention that the nonzero value of the anomalous function $g_{anom}^{R}$, strictly speaking, requires the influence of an external field of special form on the system under consideration. As a result, taking into account (1) and (41), relation (40) takes the form
\begin{eqnarray}
\left(\hbar z - \varepsilon_q\right)\,g^{R}(q,z)=1+n_0
u(q)\left\{g^{R}(q,z)+g_{anom}^{R}(q, z)\right\}. \label{C42}
\end{eqnarray}

Then we use the equation of motion for the anomalous function $g_{anom}^{R}(q, z)$ at $q\neq 0$ with Hamiltonian (7),
\begin{eqnarray}
\left(\hbar z + \varepsilon_q\right)\,g_{anom}^{R}(q,z)=-\frac{1}{V}\sum_{k\neq 0}\sum_p
u(k)\langle\langle \, a^{+}_{\bf q-k} a^{+}_{\bf p-k} a_p \mid a^{+}_{\bf q}
\rangle\rangle_{z}, \label{C43}
\end{eqnarray}
Now, as in the above consideration, we separate terms with a maximum number of the operators $a^{+}_0$ and $a_0$ on the right-hand side of (43). Then, from (43), we obtain
\begin{eqnarray}
\left(\hbar z + \varepsilon_q\right)\,g_{anom}^{R}(q,z)=-\frac{u(\bf q)}{V}
\left\{\langle\langle \, a^{+}_0 a^{+}_{-\bf q} a_0 \mid a^{+}_{\bf q}
\rangle\rangle_{z}+\langle\langle \,  a^{+}_0 a^{+}_0 a_{\bf q}\mid a^{+}_{\bf q}
\rangle\rangle_{z}\right\}, \label{C44}
\end{eqnarray}
Using the assumption (1) on the C-number representation of the particle creation $a^{+}_0$ and annihilation $a_0$ operators at a zero momentum in the calculation of Green's functions on the right-hand side of (44), we find
\begin{eqnarray}
\left(\hbar z +\varepsilon_q\right)\,g_{anom}^{R}(q,z)=-n_0
u(q)\left\{g^{R}(q,z)+g_{anom}^{R}(q, z)\right\}. \label{C45}
\end{eqnarray}
It is evident that relations (42) and (45) form a set of algebraic equations in the functions $g^{R}(q, z)$ and $g_{anom}^{R}(q, z)$, from which it immediately follows that
\begin{eqnarray}
g^{R}(q,z)=\frac{\hbar z +\varepsilon_q+n_0
u(q)}{(\hbar z)^2-(\hbar\omega(q))^2}. \label{C46}
\end{eqnarray}
where the spectrum $\omega(q)$ is defined by relation (33). Thus, the poles of the one-particle Green's function $g^{R}(q, z)$ when using the assumption (1) about the C-number representation of the particle creation $a^{+}_0$ and annihilation $a_0$ operators at a zero momentum coincide with the poles of the "density-density" Green's function $\chi^{R}(q, z)$ (see (34)).

The above consideration is based on the assumption that it is sufficient to consider the terms with a maximum number of the operators $a^{+}_0$ and $a_0$ in the calculation of Green's functions at the temperature close to zero. In this case, it is supposed that the interparticle interaction is weak.

Let us now show that the result (46) can be obtained based on the consideration of the terms with a maximum number of the operators $a^{+}_0$ and $a_0$ without the use of anomalous Green's functions $g_{anom}^{R}(q, z)$. To this end, we write the equation of motion for the function $\langle\langle \, a^{+}_{-\bf q} a_0 a_0 \mid a^{+}_{\bf q}\rangle\rangle_{z}$ (see (40)),
\begin{eqnarray}
\left(\hbar z + \varepsilon_q\right)\,\langle\langle \, a^{+}_{-\bf q} a_0 a_0 \mid a^{+}_{\bf q}\rangle\rangle_{z}=-\frac{1}{V}\sum_{k \neq 0}\sum_p
u(k)\langle\langle \left\{ a^{+}_{\bf k-q} a^{+}_{\bf p-k} a_p a_0 a_0 - a^{+}_{-\bf q} a^{+}_{\bf p-k} a_p a_{-\bf k} a_0-a^{+}_{\bf k-q} a_0 a^{+}_{\bf p-k} a_p a_{-\bf k}\right\} \mid a^{+}_{\bf q}
\rangle\rangle_{z}, \label{C47}
\end{eqnarray}
Then, on the right-hand side of (47), we separate the terms with a maximum number of the operators $a^{+}_0$ and $a_0$,
\begin{eqnarray}
\left(\hbar z + \varepsilon_q\right)\,\langle\langle \, a^{+}_{-\bf q} a_0 a_0 \mid a^{+}_{\bf q}\rangle\rangle_{z}=-\frac{u(q)}{V}
\left\{\langle\langle \, a^{+}_0 a^{+}_0 a_q a_0 a_0 \mid a^{+}_{\bf q}\rangle\rangle_{z}+a^{+}_0 a^{+}_{\bf -q} a_0 a_0 a_0\mid a^{+}_{\bf q}
\rangle\rangle_{z}\right\}, \label{C48}
\end{eqnarray}
We now use the previously advanced assumption (29) that the operator of the number of "condensate" particles $N_0=a^{+}_0 a_0$ is the C-number, rather than the operators $a^{+}_0$ and $a_0$. Then, from (48), it immediately follows that
\begin{eqnarray}
\left(\hbar z + \varepsilon_q\right)\,\langle\langle \, a^{+}_{-\bf q} a_0 a_0 \mid a^{+}_{\bf q}\rangle\rangle_{z}=-n_0 u(q)
\left\{<N_0> g^{R}(q,z)+\langle\langle \, a^{+}_{-\bf q} a_0 a_0 \mid a^{+}_{\bf q}
\rangle\rangle_{z}\right\}, \label{C49}
\end{eqnarray}
or
\begin{eqnarray}
\langle\langle \, a^{+}_{-\bf q} a_0 a_0 \mid a^{+}_{\bf q}\rangle\rangle_{z}=-\frac{n_0 u(q)}{(\hbar z)^2+\epsilon(q)+n_0 u(q)}
<N_0> g^{R}(q,z), \label{C50}
\end{eqnarray}
\textbf{If we substitute relation (50) into (40), we again obtain expression for the one-particle Green's function $g^{R}(q, z)$.}

Thus, to obtain the known results for the spectra of excitations in the degenerate Bose gas, there is no need for the assumption (1) about the C-number representation of the particle creation $a^{+}_0$ and annihilation $a_0$ operators at a zero momentum and no need for putting into consideration anomalous Green's functions (41).

Let us now return to relation (34) for the "density-density" Green's function $\chi^{R}(q, z)$, from which it follows the necessity of calculating the one-particle distribution function $f_q^T$ for "overcondensate" states. Substituting the obtained expression (46) for the one-particle Green's function into (38), we find
\begin{eqnarray}
f_q^T=\frac{\epsilon(q)+n_0 u(q)}{2\hbar \omega(q)} \coth\left(\frac{\hbar \omega(q)}{2T}\right)-\frac{1}{2}. \label{C51}
\end{eqnarray}
Let us pay attention that the dependence of the one-particle distribution function $f_q^T$ for "overcondensate" states on the order of limit transitions $T\rightarrow0$ and $q\rightarrow 0$ (see (36)) again follows from (51). At the same time, to solve the problem of the braced term $f_q^T/ \langle N_0\rangle$ in relation (34), we can reason as follows. The minimum nonzero wave vector $q$ in a specified large but finite volume $V$ is proportional to $V^{-1/3}$. Therefore, in the thermodynamic limit $V\rightarrow\infty$, $N\rightarrow\infty$, $N/V=const$, the term $f_q^T / \langle N_0\rangle$ can be considered to be zero provided that $n_0=\langle N_0\rangle/V \neq 0$, independently of the order of the limit transitions $T\rightarrow0$ and $q\rightarrow 0$.

At first sight, the above results in the limit of the weak interaction completely solve the problem of the description of the weakly non-ideal degenerate Bose gas. However, it should be taken into account that, as follows from (23), (50), the number of particles in "overcondensate" states
\begin{eqnarray}
\langle N_T \rangle= N- \langle N_0\rangle= V \int d^3 q f_q^T \label{C52}
\end{eqnarray}
is nonzero even at zero temperature. Thus, in the limit $T\rightarrow0$, the number of particles in the condensate $\langle N_0\rangle\neq N$. Hence, the assumption about the C-number behavior of the operator $N_0=a^{+}_0 a_0$, and even more so the operators $a^{+}_0$ and $a_0$ separately cannot be considered as grounded even in the limit of the weak interaction.

\section{Gap and the self-consistent distribution function}

In this regard, let us return to the consideration of equation (40) under the assumption that we correctly considered the effects of the weak interparticle interaction. Then, taking into account the above consideration, we will not assume the C-number behavior of the operator $N_0=a^{+}_0 a_0$ and the operators $a^{+}_0$ and $ a_0$ separately. Then, from the viewpoint of the perturbation theory on the interparticle interaction, the second braced term on the right-hand side of (40) is the higher-order term in comparison with the first term; therefore, it can be neglected in the assumption of our interest. Let us pay attention that, from this point of view, the result (34) for the "density-density" Green's function remains valid.

Then, in the limit of a weak interparticle interaction, from Eq.~(\ref{C40}) we obtain
\begin{eqnarray}
g^{R}(q,z)=\frac{1}{\hbar z -E_q}, \label{C53}
\end{eqnarray}
The expression for the spectrum of the one-particle excitations
$E_q$ is given by
\begin{eqnarray}
E_q=\varepsilon_q+n_0 u (q), \label{C54}
\end{eqnarray}
From Eqs.~(\ref{C38}),(\ref{C52}),(\ref{C53}), for the limit of strong degeneracy $T \ll T_0$, for the distribution function of the overcondensate particles, we
obtain
\begin{eqnarray}
f_q^T=\frac{1}{\exp(E_q/T)-1}. \label{C55}.
\end{eqnarray}
Therefore, the function $f_q^T$ is finite for $q \rightarrow 0$.
Moreover, in the limit of strong degeneracy $T \rightarrow 0$
\begin{eqnarray}
f_q^T \rightarrow 0 \label{C56}
\end{eqnarray}
for arbitrary values of $q$, in contrast to the case of the
ideal Bose gas. Therefore, the representation (\ref{C22}) for the
one-particle distribution function $f_p$ is valid, the initial
suggestions (\ref{C2}),(\ref{C18}) are satisfied and the used
procedure of extraction of the main terms is self-consistent. As is easily seen from (\ref{C42}),(\ref{C43})
for $T\rightarrow 0$ all particles in the considered approximation placed in the "condensate" with the momentum $p=0$, in contrast to
the Bogolybov's theory of weakly non-ideal homogeneous Bose gas (the effects of inhomogeneity will be considered in further publication). According to the accepted point of view, interaction between particles
leads to the quantum depletion of the condensate, which means the appearance of the particles with $p\neq 0$ at $T=0$. However, this statement is based on the assumption about coincidence of the excitation spectra for the one-particle Green function and dynamic structure factor. Only in this case the condensate density $n_0(T)$ can be calculated, based on the experimental data for $S(q,\omega)$ and the corresponding collective excitations in the spirit of the papers [22-24]. This coincidence takes place in the standard theory of weakly non-ideal Bose gas, which is constructed on the assumption about anomalous averages. This assumption is extended to liquid He II, where the approximate scheme for calculating $n_0$ is used.  Therefore, the depletion of the Bose condensate is not an experimental fact, but is the consequence of applying  of the theoretical concept of the identity of one-particle and collective spectra, obtained in the anomalous averages approach for weakly non-ideal Bose gas. It was shown above that \emph{the theories based on C-number approximations  for operators in the state with zero momentum are not self-consistent}.

In the developed theory the depletion is absent and, according to the above argumentation, there is no contradiction between this result
and the existing experiments for the dynamical structure factor, since the latter cannot be applied to calculate of the particle distribution. In general, in a more elaborated theoretical approach the depletion can appear, however, even for such approach, the coincidence of the poles of one-particle Green function and the dynamical structure factor seems an exceptional occasion. Therefore, the Bose particle distribution should be found independently of the experiments on neutron scattering.

For the \emph{self-consistent} approach under consideration the gap in the spectrum of one-particle
excitations appears and is given by
\begin{eqnarray}
\Delta=E_{q\rightarrow 0}=n_0 u(0) \label{C57}.
\end{eqnarray}
A value of the gap is completely defined by the density of particles in
the "condensate". The existence of the gap permits to extend
essentially the applicability of the results obtained for
$T\rightarrow 0$.  It is obvious, that in many applications the
condition $T\rightarrow 0$ is equivalent to the condition $T\ll
\Delta$.

Taking into account (\ref{C44}) we can rewrite Eq.(\ref{C56}) for
the function $\chi^{(0)}(q,z)$ in the form
\begin{eqnarray}
\chi^{(0)}(q,z) = \frac{2 n_0 \varepsilon_q}{(\hbar
z)^2-(\hbar\omega(q))^2}\label{C58}
\end{eqnarray}

On the basis of (\ref{C58}) almost all known results for the
thermodynamical functions of the degenerate Bose gas can be
reproduced, as is done in [9-11].

Therefore, in contrast with the approach, based on the C-number
approximation for the operators $a_0^{+}$ and $a_0$ (or applying the
C-number approximation for the operator $N_0$ in the direct perturbation theory in the presence of the condensate) we find that the spectra
of the collective and one-particle excitations are different. Both
spectra, as is easily seen, satisfy the Landau condition for
superfluidity. For the one-particle spectrum the Landau condition
is satisfied for the transitions between the "condensate"\ and the
"overcondensate"\ state. The Landau condition is, naturally,
violated for transitions between the "overcondensate"\ states.

\section{Conclusions}

Summarizing the above consideration, we can assert that the calculation of
the Green's functions for the highly degenerate Bose gas on the basis of the
C-number approximation for the operator $N_0$ and the straightforward perturbation theory allows the following
conclusions.

(i) The problem of $1/q^2$ divergence, which arises for the ideal Bose gas,
can be solved.

(ii) The system has two different branches of excitations, i.e., the
single-particle and collective ones, both satisfying the Landau condition of
superfluidity.

(iii) The single-particle excitation spectrum contains the gap in the region
of small wave vectors, associated with the existence of the \,"condensate".

(iv) The spectrum of collective excitations corresponds to "phonon-roton"
excitations observed in the experiments on inelastic neutron scattering [23,24,32].
This spectrum is the gapless mode which has no connection with the Goldstone
theorem.

(v) There is no need for anomalous averages (quasi-averages) to describe the
degenerate Bose gas.

In this connection, we emphasize that the Goldstone theorem [25]
which justifies the gapless branch of excitations is itself the
result of the breaking symmetry assumption. Therefore, it cannot
be applied to the theory considered above, which is constructed
without breaking symmetry assumption (e.g., without anomalous
averages). In our paper, we showed that the breaking symmetry is
not necessary (in this connection see also [26-29], where the
various approaches to avoid the breaking symmetry have been
developed) and at the same time the Bogolyubov result for the
collective phonon-roton mode is completely reproduced. In [30,31],
the analysis of the applicability of the relativistic Goldstone
theorem to the non-relativistic statistical theory has been done.

It should also be mentioned that Hugenholtz and Pines, constructing the
diagram technique for the degenerated Bose gas in [3], reformulated the
problem at the beginning, by changing, according to Bogolyubov, the
operators $a_0^{+}$ and $a_0$ by C-numbers. On this way, they could apply
almost automatically the quantum field theory methods to the study of the
Bose gas with \,"condensate". In particular, it was shown that, for the
C-number representation of the operators $a_0^{+}$ and $a_0$, the gap in
excitations associated with the one-particle Green's function cannot exist.
This result is also associated with the anomalous averages assumption.

In the case at hand, it should be noted that this assumption has no any
fundamental basis, is not used for the ideal Bose gas and, as we showed
above, can be avoided for the non-ideal Bose gas. Historically, the
assumption of breaking symmetry in the Bose gas theory played a crucial role
and was very useful, but in fact it is not necessary.

The proposed theory does not contradict the fundamental physical laws; the
known experiments on the collective mode can be qualitatively described
since this mode is the same as that in the existing theory based on the
breaking symmetry assumption. The Landau criterion of superfluidity is also
fulfilled.

Therefore, the interparticle interaction in Bose systems leads not only to
the drastic difference (in comparison with the ideal Bose gas) in the
structure of collective excitations, which are described by the
"density-density" Green's function, but also to the crucial change in the
distribution function of single-particle excitations and in the
single-particle excitation spectrum of for "overcodensate" particles.

Based on the results obtained, the special diagram technique can be
developed, similar to [19], but with the use of the C-number approximation
for the operator $N_0$.

The principal difference between the results of this study and the
results of the \,"traditional"\, C-number approximation for the
operators $a_0^{+}$ and $a_0$ (as well as for the operator $N_0$) is the existence of the gap in the
spectrum of single-particle excitations. The above analysis shows
that this gap cannot manifest itself in the experiments on
inelastic neutron scattering in superfluid helium (see, e.g.,
[32-34]). However, such possibility cannot be excluded [16] in the
experiments on Raman light scattering. Moreover, in [35-38], where
such experiments are described, there is a direct indication of
the existence of the gap.

Probably, the gap for single-particle excitations can manifest
itself in the experiments on the density profile in trapped Bose
gases [39]; however, this question cannot be considered within the
present paper and has to be studied separately.

\section*{Acknowledgment}
The authors thank Yu. A. Kuharenko and A.G.
Zagorodny for the helpful discussions. The authors express
gratitude to the Netherlands Organization for Scientific Research
(NWO) for support of their investigations on the problems of
statistical physics.

\end{document}